\documentclass[pmlr,twocolumn,10pt]{jmlr} 

\mlhtrack{findings}

\newif\iffinal
\finaltrue  

\iffinal
    \ifmlhneedspmlr
      \jmlrvolume{XXX}
      \jmlryear{2026}
    \fi
    \ifmlhfindings \jmlrproceedings{}{ML4H 2026 - Findings Track}\fi
    \ifmlhdemo     \jmlrproceedings{}{ML4H 2026 - Demo Track}\fi
    \jmlrworkshop{Machine Learning for Health (ML4H) 2026}
\else
    \jmlrworkshop{Machine Learning for Health (ML4H) 2026}
\fi

\usepackage{booktabs}
\usepackage{siunitx}

\usepackage[switch]{lineno}

\newcommand{\equal}[1]{{\hypersetup{linkcolor=black}\thanks{#1}}}

\newcommand{\pmsd}[1]{{\scriptsize$\,\pm$#1}}

\usepackage{booktabs} 

\usepackage{array}
\theorembodyfont{\upshape}
\theoremheaderfont{\scshape}
\theorempostheader{:}
\theoremsep{\newline}

\title[AVA: Agentic Variant Adjudication for Cancer Diagnostic Sequencing]{AVA: Agentic Variant Adjudication for Automated, Auditable Read-Level Review in Cancer Diagnostic Sequencing}

\author{%
\Name{Caleb Chin}\equal{These authors contributed equally} \Email{caleb.chin@mail.utoronto.ca}\\
\addr University of Toronto, Canada \\
\addr Centre for Addiction and Mental Health, Canada
\AND
\Name{Sidharth Gupta}\footnotemark[1] \Email{sidharth.gupta@sickkids.ca}\\
\addr University of Toronto, Canada\\
\addr The Hospital for Sick Children, Canada
\AND
\Name{Arun K {Anguraj Vadivel}}\footnotemark[1]
\Email{arun.angurajvadivel@sickkids.ca}\\
\addr The Hospital for Sick Children, Canada
\AND
\Name{Robert Siddaway}
\Email{robert.siddaway@sickkids.ca}\\
\addr The Hospital for Sick Children, Canada \\
Toronto, Canada.
\AND
\Name{Siddharth Arya}
\Email{sarya@cs.toronto.edu}\\
\addr University of Toronto, Canada
\AND
\Name{Arianna He}
\Email{arianna.he@mail.utoronto.ca}\\
\addr University of Toronto, Canada
\AND
\Name{Peter Zhukovsky}
\Email{Peter.Zhukovsky@camh.ca}\\
\addr Brain Health Imaging Centre, Centre for Addiction and Mental Health, Canada \\
\addr Department of Psychiatry, University of
Toronto, Canada.
\AND
\Name{Cynthia Hawkins}
\Email{
    Cynthia.hawkins@sickkids.ca }\\
\addr Department of Laboratory Medicine and Pathobiology, University of Toronto, Canada \\
\addr Division of Pathology, Department of Paediatric Laboratory Medicine, The Hospital for Sick Children, Canada \\
\addr The Arthur and Sonia Labatt Brain Tumour Research Centre, The Hospital for Sick Children, Canada
}

\begin{document}

\maketitle

\ifmlhdemo\else
\begin{abstract}
Identifying a tumor's DNA mutations is central to cancer diagnosis and treatment, but distinguishing genuine mutations from sequencing artifacts is error-prone. In practice, genuine mutations are identified using fixed variant-allele-frequency (VAF) and depth thresholds which cannot separate artifacts from genuine mutations in the borderline low-VAF zone where they overlap. A laboratory specialist must manually inspect the read-level sequencing data, a slow step that has high inter-rater variability and causes diagnosis delays. We present AVA: Agentic Variant Adjudication; an agentic framework that automates this read-level inspection with subagents that reason over evidence, generate investigative code, and return a real-or-artifact adjudications with an auditable reasoning trace all without training on existing data. On both the SEQC2 DNA-seq benchmark and patient RNA-seq data, restricted to a subsample of borderline variants where thresholds fail, both closed-source (Claude Sonnet 5) and open-weight (GLM 5.3 Flash and Gemma 4 31b) agents outperform the best achievable VAF/depth thresholds. Ultimately, AVA is suited for privacy-constrained clinical deployment, since the framework can run on locally deployed open-weight models with descriptive reasoning traces to keep specialists in the loop.

\end{abstract}
\fi

\ifmlhneedsstatements
\paragraph*{Data and Code Availability}

 We release our full framework: the orchestrator \& sub-agent prompt templates, the read-level feature extractor, all data analysis \& evaluation scripts together with the complete per-variant reasoning traces, at this \href{https://github.com/calebchin/AVA}{GitHub repository}. Experiments use the publicly available SEQC2 benchmark (\cite{fang2021establishing}), obtained from the NCBI SEQC2 repository, using the whole-genome tumor/normal alignments and the high-confidence variant truth set. Additionally, experiments evaluate on RNA-seq data from \cite{siddaway2025} available at accession EGAS50000000700.


\paragraph*{Institutional Review Board (IRB)}
This study used de-identified clinical RNA-seq data from The Hospital for Sick Children (SickKids), previously described by \cite{siddaway2025}. This work was conducted under approval from The Hospital for Sick Children Research Ethics Board (REB 1000055059). All analyses were performed locally.
\fi

\section{Introduction}
\label{sec:intro}

Many cancers, such as lung cancers or gliomas, are genomically characterized. For example, gliomas are further subtyped into categories such as astrocytomas or glioblastomas based on mutations present in the tumor DNA (\cite{louis20212021}). When a patient is suspected of having a tumor, a biopsy is commonly performed, which is a medical procedure that samples biological material (e.g. blood, cerebral spinal fluid, or tumor tissue) from the patient for diagnostic testing. Genomic sequencing may be run on the patient’s biopsy sample to detect mutations in the tumor DNA, which are then used by the oncology team to form the patient’s diagnosis and treatment plan.

Genomic sequencing of the tumor biopsy produces millions of short DNA reads, which are aligned to a reference genome to form a stack (called a pileup) of overlapping tumor DNA reads at each reference genomic position (locus). A locus where some reads in the pileup disagree with the reference (e.g., an A instead of a T) is called a variant. A variant describes a possible mutation, but still must be distinguished from a technical sequencing artifact. Each variant is summarized by its depth (the number of reads covering the locus) and its variant allele frequency (VAF, the fraction of reads in the locus that disagree with the reference and support this particular variant). In practice, which reported variants are called mutations is often governed by simple cutoffs on VAF and depth (\cite{Cheng2015MSKIMPACT}, \cite{jennings2017guidelines}, \cite{siddaway2025}). For example, a variant may be classified as a true mutation only when its VAF exceeds 10\% and its depth exceeds 20 reads.

Many therapies (immunotherapies, targeted therapies, etc) are designed specifically to target specific mutations (\cite{mateo2022delivering}), and will most benefit patients whose tumors present those mutations. Thus, in cancer diagnosis we must identify all of a patient’s tumor DNA mutations as completely and accurately as possible, while limiting delays in diagnosis.
However, the process of going from the patient biopsy sample to the final list of identified mutations is error-prone, and mutations that are missed or incorrectly identified can delay diagnosis and treatment planning. One significant source of error is that there is no optimal threshold for VAF and depth to classify variants as genuine mutations: any fixed cutoff necessarily misclassifies some true mutations as artifacts and some artifacts as true mutations in an overlapping zone we term the ``borderline'' zone (Figure \ref{fig:borderlinezone}). 

\begin{figure}[t]
\floatconts
{fig:borderlinezone}
  {\caption{VAF borderline zone (left) and threshold cutoff (right) showing true mutations \&  artifacts overlap}}
  { \includegraphics[width=0.4335\textwidth]{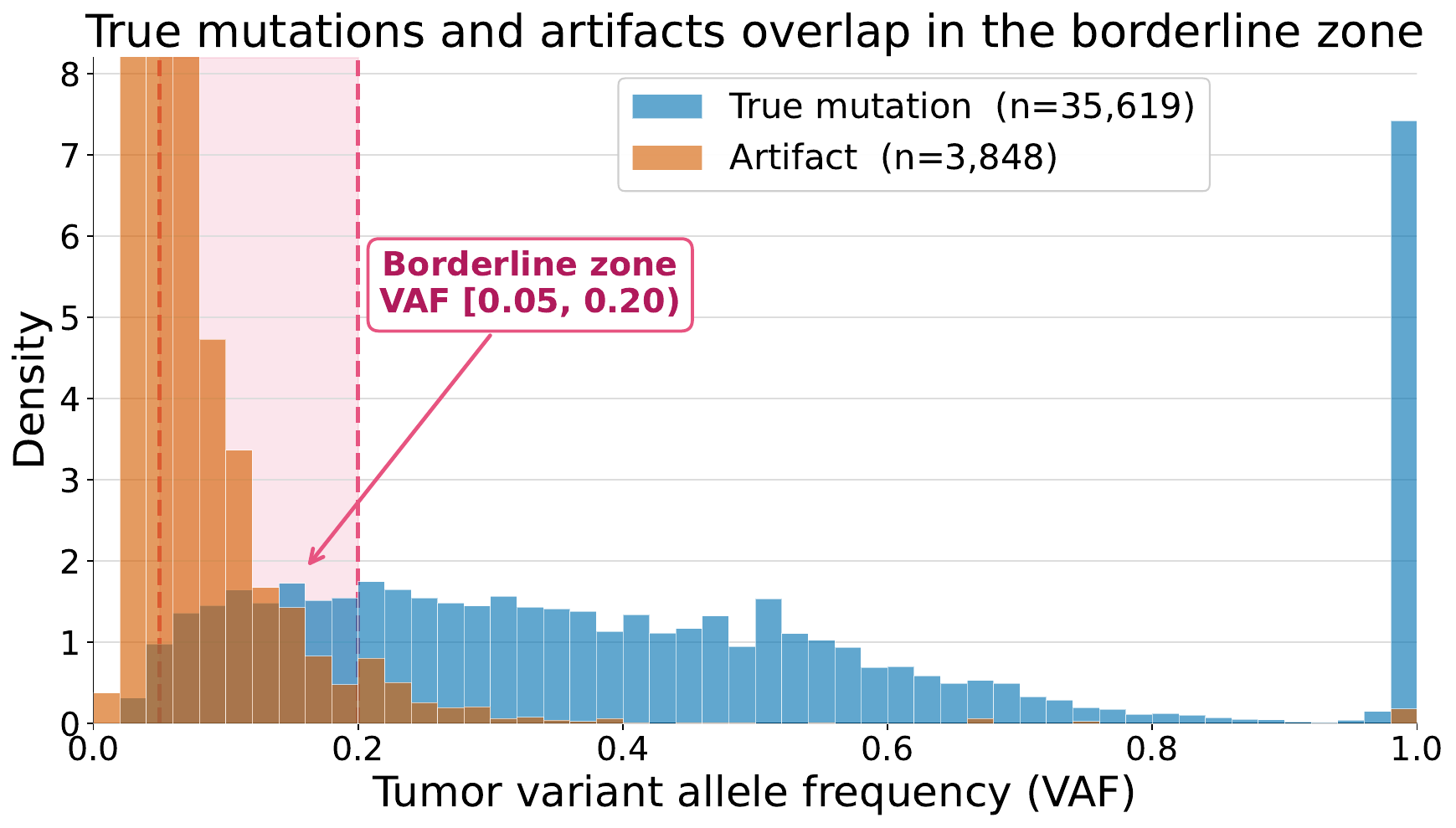}
    \vspace{-20pt}
    }
    \vspace{-16pt}
\end{figure}
\begin{figure*}[t]
    \centering
    \includegraphics[width=1.0\linewidth]{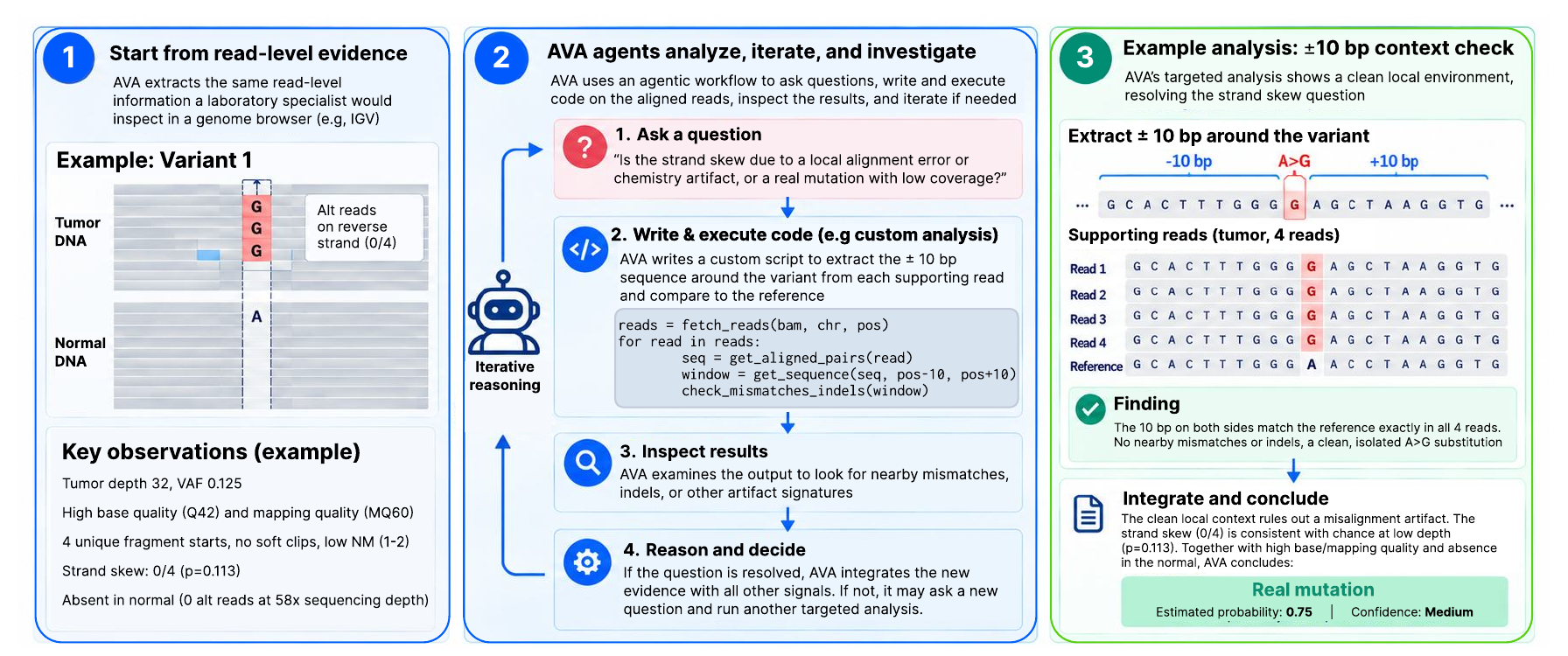}
    \caption{The AVA workflow, illustrated on Variant 1. AVA first extracts the same read-level evidence a laboratory specialist would inspect in a genome browser. When this initial evidence is ambiguous, AVA autonomously formulates a targeted question, writes and runs custom code against the aligned sequencing reads, inspects the result, and iterates as needed. For Variant 1, the one worrying signal is that all four variant-supporting reads fall on the same DNA strand, which is a possible artifact. This prompts AVA to check if those reads are misaligned. Its custom analysis of the sequence around the variant finds no nearby errors, ruling out misalignment, and AVA calls the variant a real mutation with medium confidence, recovering a true variant that a fixed threshold discards.}
    \label{fig:ava_workflow}
\end{figure*}
Variants in a borderline zone must be inspected manually by a laboratory specialist, who examines read-level features to distinguish genuine mutations from sequencing artifacts. For a given case, a specialist will typically open the variant in a genome browser such as IGV (\cite{robinson2011integrative}) and analyze the pileup of individual reads covering the locus. From this pileup, they compute and inspect a set of features (examples listed in Table \ref{tab:read-features}), and from these features they judge whether the candidate reflects a genuine mutation or a technical artifact. Each of these features are informative because they expose a distinct way the sequencing pipeline can fail, which is described in more detail in Table \ref{tab:read-features}. In this task of manually interpreting the detected variants, \cite{sirohi2020multi} found that laboratory specialists and pathologists across institutions have significant inter-rater variability, indicating that manual review can both be inefficient and variable. 

To address this, we propose AVA (Agentic Variant Adjudication), an agentic framework that resolves the ambiguity of fixed VAF/depth thresholds by emulating a specialist's read-level review. In this framework, an orchestrator launches one sub-agent per borderline variant into a code-execution workspace. Each sub-agent examines its variant pileup, autonomously queries the aligned sequencing data to gather additional evidence if needed, and ultimately classifies the variant as a true mutation or a technical artifact (Figure \ref{fig:ava_workflow}). As every verdict includes an auditable justification citing the specific features behind the call, AVA's reasoning remains transparent. We evaluate AVA on variants in this borderline threshold zone (Figure \ref{fig:borderlinezone}). 

Benchmarking the performance of open-weight models against frontier cloud models is crucial for demonstrating clinical utility as patient genomic data often cannot be transmitted to third-party cloud services for privacy, consent, and data-residency reasons. Therefore, on the public SEQC2 dataset, we evaluate AVA with both a frontier cloud model (Claude Sonnet 5 \cite{anthropic2026claudesonnet5}) and locally-deployable open-weight models of various sizes (GLM 5.3 Flash \cite{glm5team2026glm5vibecodingagentic} and Gemma 4-31b \cite{gemmateam2026gemma4technicalreport}). 

We validate on SEQC2, a publicly available cell-line dataset with high-confidence ground-truth labels for true mutation and artifact variants. We also validate on a controlled-access RNA-seq patient dataset from \cite{siddaway2025} using locally deployed models. Our findings show that AVA models exceed the best achievable VAF/depth threshold baseline, and we further qualitatively probe these findings by examining the reasoning traces of the agents in our experiments.

\begin{table*}[t]
\centering\small
\setlength{\tabcolsep}{4pt}
\caption{Performance on 300 borderline SEQC2 variants (150 true, 150 artifact). LLM rows report mean\,$\pm$\,s.d.\ over 5 runs; FN and FP are counts. EC-Acc.\ is accuracy on 40 easy-control variants. Threshold rules are tuned on the evaluation labels (optimistic upper bound). $^\dagger$SEQC2 labels were partly derived from FilterMutectCalls, so this row is also an optimistic upper bound.}
\label{tab:seqc2}
\begin{tabular}{@{}lccccccc@{}}
\toprule
Method & Acc. & Prec. & Rec. & F1 & FN & FP & EC-Acc. \\
\midrule
\multicolumn{8}{@{}l}{\emph{AVA (agentic)}}\\
\quad Claude Sonnet 5 & .903\pmsd{.029} & .866\pmsd{.049} & .956\pmsd{.030} & .908\pmsd{.027} & 6.6\pmsd{4.4} & 22.4\pmsd{8.7} & .910\pmsd{.034}\\
\quad Gemma 4 31b     & .921\pmsd{.018} & \textbf{.922\pmsd{.013} }& .920\pmsd{.044} & .920\pmsd{.020} & 12.0\pmsd{6.6} & \textbf{11.6\pmsd{2.3}} & .955\pmsd{.011}\\
\quad GLM 5.3 Flash   & \textbf{.927\pmsd{.008}} & .881\pmsd{.013} & \textbf{.988\pmsd{.003}} & \textbf{.931\pmsd{.007} }& \textbf{1.8\pmsd{0.4}} & 20.0\pmsd{2.6} & .930\pmsd{.011}\\
\midrule
\multicolumn{8}{@{}l}{\emph{Baselines}}\\
\quad GLM 5.3 Flash (non-agentic) & .900\pmsd{.009} & .840\pmsd{.014} & .988\pmsd{.006} & .908\pmsd{.007} & 1.8\pmsd{0.8} & 28.2\pmsd{3.0} & .920\pmsd{.011}\\
\quad FilterMutectCalls$^\dagger$ & .920 & .920 & .920 & .920 & 12 & 12 & .975\\
\quad Best threshold (VAF$\times$depth\,$\geq$\,7) & .810 & .855 & .747 & .797 & 38 & 17 & \textbf{1.000}\\
\bottomrule
\end{tabular}
\end{table*}

\section{Related work}


\cite{barnell2019standard} formalized the manual variant adjudication process into a standard operating procedure, and showed that this standardized process improves reviewer accuracy and reduces (but does not eliminate) inter-reviewer variability. \cite{ainscough2018deep} took the first large-scale step toward automating this process, training random forest and deep learning classifiers on ${\sim}$41,000 manually reviewed variants (an estimated 585 hours of expert effort) over 71 fixed, predominantly read-level features. These models matched held-out manual labels (AUC 0.96-0.98) and orthogonal validation sequencing (AUC 0.95-0.96), but as supervised models they are tied to the data they were trained on: on external cohorts with different sequencing and experimental conditions, performance degraded, most severely for ambiguous calls (AUC 0.43-0.47). Restoring performance required retraining on newly labeled variants, and the authors recommend manually reviewing a subset (e.g., 5\%) of every new dataset for this purpose, so each deployment still carries a site-specific labeling cost. Large language models (LLMs) offer a potential route around this labeling dependency: \cite{lin2025benchmarking} showed that general-purpose LLMs can interpret cancer variants by clinical actionability without task-specific training. To our knowledge, artifact adjudication has not benefited from this shift: beyond fixed filters and thresholds, automated approaches remain mainly supervised. AVA closes this gap with a zero-shot adjudicator, an agentic framework that analyzes read-level evidence from the aligned sequencing data and requires no labeled variants or site-specific retraining.

\section{Dataset and experimental design}

We use SEQC2, a reference dataset built from a matched tumor/normal cell-line pair sequenced by whole-genome sequencing (Illumina, GRCh38) (\cite{fang2021establishing}). Variants are the single-nucleotide calls emitted by a somatic variant caller (MuTect2, \cite{benjamin2019calling}) run on the tumor/normal pair. Ground truth comes from the SEQC2 high-confidence single-nucleotide variant set, a consensus list of true variants for this cell line that the consortium established by cross-validating calls across many sequencing centers, sequencers, aligners, and callers, and which serves as the community gold standard.

On SEQC2, we define the borderline zone as the VAF interval [0.05, 0.20), where both true mutation and artifact classes are well represented and no VAF or depth cutoff cleanly separates them. We draw a balanced evaluation set of 340 variants. The main benchmark is the borderline set: 300 variants sampled from the borderline zone, with 150 true mutations and 150 artifacts. As a sanity check, we add an easy control set of 40 variants drawn far outside the overlap, 20 unambiguous true calls at high VAF and 20 unambiguous artifacts at very low VAF, on which thresholding methods should perform near-perfectly.
Before running any agent, we sweep VAF, depth, and their combination (VAF multiplied by depth) to find the highest accuracy any such threshold rule attains on the borderline set. The resulting rule is an optimistic upper bound baseline for thresholds. 
We also use the FilterMutectCalls package from the Genome Analysis Toolkit (GATK) (\cite{vanderauwera2020genomics}) as an additional baseline. FilterMutectCalls utilizes a probabilistic model over a larger feature set than just VAF and depth. An important caveat is that FilterMutectCalls was one of the callers used to label the SEQC2 data, so it too represents an optimistic upper bound on baseline performance. 

Additionally, we take patient RNA-seq data from \cite{siddaway2025}'s research study. In this study, a clinical threshold is described -- VAF $>$ 10\%, depth $\geq$ 20 and at least 5 alt reads (VAF $\times$ depth $\geq$ 5). We define two evaluation bins, one is below this clinical threshold, and one is borderline (passes the clinical threshold, with 10\% $\leq$ VAF $\leq$  20\%). \cite{siddaway2025} report the final list of true variants in their supplementary materials, and from this list, we sample 13 true mutation variants from the below threshold bin, and 15 true mutation variants from the borderline bin. We construct proxy artifact variants by applying 9 filters described in Appendix Table \ref{tab:artifact-filters}. These filters flag variants that exist in the RNA-seq data and signal reproducible errors, non-RNA editing variants, and variants that are unlikely to represent tumor biology. We sample 13 proxy artifact variants from the below threshold bin, and 15 proxy artifact variants from the borderline bin. We utilize AVA with the Gemma 4 31b model to ensure patient data is kept local and private. Like our experiments on SEQC2, we evaluate on a threshold optimally tuned on these variants, as well as the original clinical threshold.


AVA utilizes the Claude Code multi-agent harness (\cite{anthropic_claude_code_2026}) to give each agent a code execution environment. An orchestrator agent launches one sub-agent per variant. Each sub-agent receives the locus (chromosome, position, reference base, alternate base) and a stream to the aligned sequencing data, with no truth label, no VAF / depth, and no indication of whether the variant is borderline. The sub-agent first runs a feature extractor, which streams the tumor and matched-normal reads at that locus from the sequencing data, and returns the read-level panel of Table \ref{tab:read-features}. Unlike a static classifier, the sub-agent can iteratively query the aligned sequencing data; if the initial evidence is insufficient, it autonomously writes and executes its own scripts to investigate further read-level features (e.g., sequence context). It reasons over this new evidence and iterates until it returns a verdict (true or artifact), a probability that the variant is real, a confidence level, and a justification citing the specific features and analyses that drove its decision. Each sub-agent runs in its own isolated workspace, and never sees any other variant or its label. Figure \ref{fig:ava_workflow} illustrates the AVA framework in detail.

To separate the utility of an agentic harness (i.e. code execution, tool calling) from the internal knowledge of the underlying LLM, we also introduce an ablation experiment. We run all adjudications through an inference-only pass of GLM 5.3 Flash using the same static features that the feature extraction script given to AVA models. These adjudications are made without a code execution environment or tool calling, but the base AVA prompt remains structurally the same. 

\begin{table*}[t]
\centering\small
\setlength{\tabcolsep}{5pt}
\caption{Performance on 56 borderline variants (28 true, 28 artifact) from the patient RNA-seq dataset \citep{siddaway2025}. AVA reports mean\,$\pm$\,s.d.\ over 3 runs; FN and FP are counts.}
\label{tab:rnaseq}
\begin{tabular}{@{}lcccccc@{}}
\toprule
Method & Acc. & Prec. & Rec. & F1 & FN & FP \\
\midrule
AVA (Gemma 4 31b) & .821\pmsd{.018} & .898\pmsd{.016} & .726\pmsd{.055} & .802\pmsd{.027} & 7.7\pmsd{1.5} & 2.3\pmsd{0.6}\\
Clinical threshold \citep{siddaway2025} & .500 & .500 & .536 & .517 & 13 & 15\\
Best threshold (VAF\,$\geq$\,0.11, depth\,$\geq$\,18) & .625 & .590 & .821 & .687 & 5 & 16\\
\bottomrule
\end{tabular}
\end{table*}


\section{Results}

\subsection{Quantitative results}

\paragraph{SEQC2 Data}
Table \ref{tab:seqc2} reports performance on the 340 variants for all models and the threshold baselines across 5 independent seeded runs. Across all 5 seeds, all AVA models exceed the VAF and depth optimistic upper-bound threshold rule in almost all reported metrics. Notably, the best performing AVA model GLM 5.3 Flash had accuracy 0.927 $\pm$ 0.008 vs the best threshold's accuracy of 0.810. AVA models also reduced the number of false negatives (FNs) compared to the threshold in this setting (from 38 FNs to 1.8 $\pm$ 0.4). We will further discuss the false negative and false positive (FPs) tradeoff in Section {\ref{sec:disc}}. Among the easy control cases, the best threshold achieves perfect classification as expected, whereas AVA performs worse on average (EC-Acc $\sim$0.932 across AVA models). However, this discrepancy can be resolved by applying a threshold to classify easy cases prior to running AVA on the borderline cases.

When compared to the FilterMutectCalls baseline, AVA models match its performance across seeds (GLM 5.3 Flash acc: 0.927 $\pm$ 0.008 vs 0.920). One important caveat is that FilterMutectCalls was one of the callers used to label the SEQC2 data, so it also represents an optimistic upper bound on performance.

We also measure the performance of the base LLMs without the code execution environment and tool calling. GLM 5.3 Flash (non-agentic) outperforms the best threshold baseline (acc 0.900 $\pm$ 0.009), but not FilterMutectCalls or its agentic GLM 5.3 Flash equivalent. Interestingly, the agentic GLM 5.3 Flash holds the FNs constant (1.8) compared to the non-agentic version, but reduces the number of FPs (28.2 to 20). 

\paragraph{RNA-Seq Patient Data}
Table \ref{tab:rnaseq} shows results for AVA across 3 seeds against the clinical threshold described by \cite{siddaway2025} and an optimally tuned VAF and depth threshold on the evaluation set. We remark that AVA consistently outperforms both the clinical and best threshold by a significant margin across all seeds (mean accuracy .821 versus 0.625 on the best threshold). In addition, AVA returns significantly fewer false positives than both the clinical and best threshold (mean 2.3 FP versus 15 and 16 on the clinical and best thresholds).

\subsection{Qualitative Results}
Two cases show why reading the individual reads beats thresholding on summary statistics. Variant 1 (chr2:37379311, A$>$G) is a real mutation with too few supporting reads to pass a count threshold, so the threshold discards it as noise (a false negative). The agent wrote and ran a short script to check the sequence around the candidate on each supporting read: a genuine read differs from the reference at the candidate position and nowhere nearby, whereas extra errors in the neighborhood would mean the read was placed at the wrong genomic location. Finding the neighborhood clean, the reads high-quality, and the allele absent from the patient's matched healthy sample, it accepted the mutation the threshold had thrown away (Figure~\ref{fig:trace-v1}). Variant 64 (chr1:13372097, C$>$T) is the mirror image: it has enough support to look real, but the sequence at this site also appears almost identically at another location in the genome, so the aligner cannot tell which of the two the reads actually came from. The agent flagged this ambiguity, noted the same allele in the healthy sample, and rejected it as an artifact invisible to VAF and depth (Figure~\ref{fig:trace-v64}). In both cases AVA reaches a specialist's judgment and logs the features behind it, so a specialist can audit or overrule the call.

Trace transparency also exposes systematic failure modes. When only a handful of reads support a candidate, AVA tends to over-weight a few clean reads as positive evidence and discount weak signals pointing to artifact. In Variant 31 (chr2:73231720, C$>$G), we see AVA dismissing one significant artifact signal (a variant read in the matched normal), and over-weighting two clean reads, producing a false positive (Figure~\ref{fig:trace-fp}). Characterizing this bias is future work.

\section{Discussion}\label{sec:disc}
Our results indicate that read-level adjudication by a large language model agent can separate true variants from artifacts in exactly the borderline zone where fixed VAF/depth thresholds fail. In this zone, both the closed-source and open-weight models exceed the upper bound best achievable VAF/depth threshold rule across accuracy and F1-score in both DNA-seq and real patient RNA-seq data. Because the agent reasons over the same read-level evidence a specialist inspects in a genome browser, and records a inspectable trace for each call, it functions as an auditable second reader, directly addressing the inter-rater variability discussed by \cite{sirohi2020multi}. We also find that Gemma 4 31b, a locally-deployable open-weight LLM, matches the performance of much larger models (GLM 5.3 Flash and Claude Sonnet 5) across several metrics. This finding is particularly significant for clinical use, where patient genomic data frequently cannot leave the institution and compute resources are limited. 

We further find that a considerable portion of the performance gains above the baseline thresholds can be recovered with an inference-only pass of the underlying LLM (see GLM 5.3 Flash (non-agentic) in \ref{tab:seqc2}), indicating that the internal LLM knowledge combined with the AVA prompting is capable of variant adjudication without the code execution environment and tool calls. However, we emphasize the utility of the agentic harness beyond the reported performance metrics. In particular, the ability to perform tool-calling and code execution allows the agent to utilize a dynamic set of features much like a laboratory specialist, which could be useful in more complex patient cases. 

Although AVA greatly reduces the number of false negatives in the SEQC2 evaluation, we see an increase in false negatives in the smaller patient RNA-seq dataset compared to threshold baselines (see Table \ref{tab:rnaseq}). Clinical sequencing workflows typically favor sensitivity to avoid missing clinically important variants, and accept higher false-positive rates that are then resolved by further clinical testing (such as immunohistochemistry) (\cite{LINCOLN2019318}). However, the additional testing adds costs and delays results. Balancing the sensitivity to important variants and the need for efficient adjudication is therefore a difficult task and may need to be tuned to institutional preferences.

\vspace{-4pt}
\paragraph{Limitations} While our work shows promise in deploying agents in real-world settings for variant adjudication, there are a few key limitations in our findings. 

First, the SEQC2 dataset is a single public breast cancer cell-line dataset which differs from real patient biopsies. For example, FFPE-preserved clinical biopsies may carry formalin-induced low-VAF biases that are absent from SEQC2's high-quality cell-line DNA (\cite{guo2022mutational}). Additionally, as SEQC2 is a popular public dataset, there is a possibility that it appeared within the training corpus of the LLMs that we have tested. 

Second, while the RNA-seq dataset addresses both concerns above, the current sample size is limited (n = 56) and we only test a single model's performance (Gemma 4:31b) due to computational constraints.  

Third, we utilize a single set of prompts and harness configurations for our AVA experiments and have not yet evaluated prompt or task environment stability, which are key milestones for generalizable clinical utility.

Finally, we observe certain failure modes in which AVA reasons in a biased way, such as in Variant 31. We have yet to characterize exactly where this bias stems from and how to combat it. 


\vspace{-4pt}
\paragraph{Future work} To address these limitations, future studies will validate AVA on larger patient aligned sequencing data within real clinical workflows across a variety of model configurations. Studying AVA's utility in real clinical settings can elucidate it's ability to provide clinical impact, e.g in reducing patient turnaround time, preventing unnecessary clinical tests with reduced false positives, etc. We also propose a more in-depth analysis of prompt stability and optimization by following an evaluation protocol similar to \cite{rabanser2026scienceaiagentreliability} through prompt ablations and environment perturbation.

Stemming from the discussion about balancing FPs and FNs, future work on AVA will explore ways to increase sensitivity while maintaining a small number of FPs on patient data, such as inspecting AVA's reasoning traces for review or having a shared knowledge bank for subagents to search from and contribute to. 

Demonstrating the early utility of a simple framework like AVA opens the door to more complex multi-agent architectures. For example, difficult variant adjudications could be viewed by a voting committee of agents. Further, AVA could be used to identify and investigate large-scale patterns of common artifacts, which can potentially be used to develop new feature sets that can better characterize real vs. artifact variants. We leave these advancements as interesting future work.

\section*{Acknowledgments}

We thank Dr.\  Chris Maddison, Dr.\  Martin Beaulieu, Dr.\  Rebecca Laposa, and the course staff of CSC2541: AI for Drug Discovery (Winter 2025) for their help in developing ideas for an early version of this project. We also thank Dr. Gary Bader and Mansuba Rana for their conversations that helped describe the variant adjudication problem space in this work.

\bibliography{ref}

\clearpage
\onecolumn
\appendix

\section{Appendix}\label{apd:first}

\begin{table}[ht]
  \centering
  \small
  \setlength{\belowcaptionskip}{6pt}
  \caption{Filters used to select proxy artifact variants from the RNA-seq evaluation dataset released in \cite{siddaway2025}
    An artifact candidate is retained only if it passes all nine filters.}
  \label{tab:artifact-filters}
  \renewcommand{\arraystretch}{1.25}
  \begin{tabular}{@{}c
      >{\raggedright\arraybackslash}p{0.52\linewidth}
      >{\raggedright\arraybackslash}p{0.40\linewidth}@{}}
    \toprule
    \textbf{\#} & \textbf{Filter} & \textbf{Purpose} \\
    \midrule
    \multicolumn{3}{@{}l}{\textit{Recurrence}} \\
    1 & $\geq$3 alt reads in $\geq$10 samples, from $\geq$5 different runs
      & Reproducible error, not chance noise or one bad batch \\
    2 & Called in $\leq$10\% of runs ($\leq$37 of 370)
      & Mirrors the pipeline's recurrence filter \\
    \addlinespace
    \multicolumn{3}{@{}l}{\textit{Germline and population frequency}} \\
    3 & Never above 30\% VAF in any sample & Not an inherited variant \\
    4 & gnomAD AF blank or $\leq$1\% & Mirrors the pipeline's gnomAD filter \\
    \addlinespace
    \multicolumn{3}{@{}l}{\textit{Known biological sources}} \\
    5 & Not A$\to$G / T$\to$C & Exclude RNA editing \\
    6 & Outside HLA/MHC, IG and TCR loci & Exclude real biological variation \\
    \addlinespace
    \multicolumn{3}{@{}l}{\textit{Clinical relevance}} \\
    7 & Gene appears in Table~S1 of \cite{siddaway2025} & Genes a specialist actually reviews \\
    8 & Missense or nonsense only & Changes that could be reported \\
    9 & Not a change or position from variants in Table S1 of \cite{siddaway2025}, nor within $\pm$1 codon of one
      & Exclude real hotspot variants \\
    \bottomrule
  \end{tabular}
\end{table}

\begin{table}[h]
\floatconts
  {tab:read-features}
  {\caption{Read-level pileup features a laboratory specialist will inspect, and the artifact mechanism each reveals}}
  {\begin{tabular}{@{}p{0.45\linewidth}p{0.45\linewidth}@{}}
    \toprule
    \textbf{Read-level feature} & \textbf{What it reveals} \\
    \midrule
    Strand balance of variant-supporting reads & Strand bias \\
    Base quality of the variant base & Low-confidence base calls \\
    Mapping quality (fraction with MAPQ 0) & Ambiguous read placement \\
    Position of the variant within reads & Clustering at read ends \\
    Per-read mismatch count and soft-clipping & Messy or clipped alignments \\
    Distinct fragment starts vs.\ duplicate stacks & PCR-duplicate inflation of apparent support \\
    Presence of the variant in the matched normal & Non-tumor-specific signal \\
    \bottomrule
  \end{tabular}}
\end{table}

\begin{figure}[h]
\caption{Reasoning trace review, Variant 1 (chr2:37379311, A$>$G), recovering a real variant the threshold discards.
  Ground truth: real variant. Model verdict: real, $p=0.75$, medium confidence (correct).}
\label{fig:trace-v1}
\vspace{0.5em}\hrule\vspace{0.5em}
{\small
\textbf{What the model saw}\par\vspace{0.3em}
\begin{tabular}{@{}p{3.0cm} p{3.2cm} p{8.6cm}@{}}
\textbf{Signal} & \textbf{Value} & \textbf{What it indicates}\\
\hline
Variant reads & 4 of 32 (VAF 0.13) & Below a typical count threshold, so a threshold rule discards it as noise\\\\
Base quality & Q42, all 4 reads & Base calls are high-confidence \hfill($\rightarrow$ real)\\\\
Mapping quality & MQ60, all 4 reads & Reads confidently placed in the genome \hfill($\rightarrow$ real)\\\\
Nearby errors on reads & none within $\pm$10\,bp & Reads differ from the reference only at the variant, so they are not misaligned \hfill($\rightarrow$ real)\\\\
Independent fragments & 4 of 4 & Distinct molecules, not PCR duplicates \hfill($\rightarrow$ real)\\\\
Strand & 0 fwd / 4 rev & All reads on one strand: a mild red flag, but expected by chance at only 4 reads \hfill(weak)\\\\
In healthy sample & 0 of 58 & Variant is tumor-specific \hfill($\rightarrow$ real)\\
\end{tabular}
\vspace{0.5em}\hrule\vspace{0.5em}
\textbf{What the model did}\par\vspace{0.3em}
A depth threshold would discard this variant since only four reads support the variant. However the model inspected the reads directly. All four are high-quality, and the variant is absent from the healthy sample; the only concern is that all four fall on one strand. To test whether that hinted at an artifact, the model wrote and ran a short script that compared each read to the reference sequence in a short window on either side of the variant. Each read matched the reference everywhere except at the variant base itself, indicating no misalignment errors.
\vspace{0.5em}\hrule\vspace{0.5em}
\textbf{Outcome}\par\vspace{0.3em}
\textbf{Correct.} Independent, high-quality, tumor-specific reads with clean surroundings outweigh the low read count and the single-strand skew, so the model recovers a real variant that the threshold rule missed.
\hrule
}
\end{figure}

\begin{figure*}[t]
\caption{Reasoning trace review, Variant 64 (chr1:13372097, C$>$T), rejecting an artifact that looks real.
  Ground truth: artifact. Model verdict: artifact, $p=0.04$, high confidence (correct).}
\label{fig:trace-v64}
\vspace{0.5em}\hrule\vspace{0.5em}
{\small
\textbf{What the model saw}\par\vspace{0.3em}
\begin{tabular}{@{}p{3.0cm} p{3.6cm} p{8.2cm}@{}}
\textbf{Signal} & \textbf{Value} & \textbf{What it indicates}\\
\hline
Variant reads & 7 of 57 (VAF 0.12) & Enough support to look real by count and VAF\\\\
Base quality & Q42 median & High, clean on the surface (looks real)\\\\
Mapping quality & MQ21 median (reference reads MQ60) & Variant reads placed with much lower confidence than reference reads at the same locus ($\rightarrow$ artifact)\\\\
Extra mismatches per read & 4 median & Variant reads carry many other differences from the reference: messy alignments ($\rightarrow$ artifact)\\\\
Alternate-alignment flag & 7/7 tumor, 2/2 healthy & Every variant read also fits another, near-identical locus $\sim$146\,kb away ($\rightarrow$ artifact)\\\\
In healthy sample & 2 of 38 & Not tumor-specific ($\rightarrow$ artifact)\\
\end{tabular}
\vspace{0.5em}\hrule\vspace{0.5em}
\textbf{What the model did}\par\vspace{0.3em}
Base quality and strand balance look clean, so the variant appears real at first. But the variant reads have far lower mapping quality than the reference reads at the same locus, a sign they may be mis-located. To check, the model wrote and ran a short script inspecting each read's alternate-alignment flags, and found that every variant read also matches a near-identical sequence about 146\,kb away. The aligner therefore cannot tell which of the two locations the reads came from, and the same allele also appears in the healthy sample.
\vspace{0.5em}\hrule\vspace{0.5em}
\textbf{Outcome}\par\vspace{0.3em}
\textbf{Correct.} The reads cannot be uniquely placed and are not tumor-specific, so the model rejects a mapping artifact that allele fraction and read depth alone cannot reveal.
\hrule
}
\end{figure*}
\begin{figure}[t]
\caption{Reasoning trace review, Variant 31 (chr2:73231720, C$>$G), a false positive that exposes the model's bias.
  Ground truth: artifact. Model verdict: real, $p=0.74$, medium confidence (incorrect).}
\label{fig:trace-fp}
\vspace{0.5em}\hrule\vspace{0.5em}
{\small
\textbf{What the model saw}\par\vspace{0.3em}
\begin{tabular}{@{}p{3.0cm} p{3.6cm} p{8.2cm}@{}}
\textbf{Signal} & \textbf{Value} & \textbf{What it indicates}\\
\hline
Variant reads & 3 of 38 (VAF 0.08) & Very few reads, inconclusive on its own\\\\
Base quality & Q32 median & Adequate \hfill(weakly $\rightarrow$ real)\\\\
Mapping quality & MQ43 median & Adequate, reads reasonably placed \hfill(weakly $\rightarrow$ real)\\\\
Nearby errors on reads & 2 of 3 clean; 1 read has extra mismatches near its end & Most reads look clean \hfill(weakly $\rightarrow$ real)\\\\
Independent fragments & 3 of 3 & Distinct molecules, not duplicates \hfill(weakly $\rightarrow$ real)\\\\
In healthy sample & 1 of 39 (low quality) & A variant read \emph{is} present in the healthy sample ($\rightarrow$ artifact)\\
\end{tabular}
\vspace{0.5em}\hrule\vspace{0.5em}
\textbf{What the model did}\par\vspace{0.3em}
With only three reads, the model inspected each one. Two looked ``pristine'', being a single mismatch at the variant, clean surroundings, on opposite strands, and AVA treated these as strong support for a real variant. It dismissed the one variant read in the healthy sample as low-quality noise, and concluded that the clean reads outweighed the low read count.
\vspace{0.5em}\hrule\vspace{0.5em}
\textbf{Outcome and interpretation}\par\vspace{0.3em}
\textbf{Incorrect (false positive; ground truth: artifact).} This trace shows the two-part bias described in the Discussion. First, AVA over-weights the two clean reads as positive evidence, though at only three reads they are equally consistent with a real mutation and with random noise. Second, it discounts the one signal pointing to artifact --- a variant read in the matched normal, evidence of a shared, non-tumor artifact --- dismissing it as low-quality noise. The contrast with Variant~1 (Figure~\ref{fig:trace-v1}) is instructive: there the same clean-reads reasoning was sound \emph{because} the matched normal was genuinely empty (0 of 58); here that corroboration is absent, yet AVA reaches the same verdict by ignoring the artifact signal.
\hrule
}
\end{figure}

\end{document}